\documentclass[twocolumn, prx, superscriptaddress, notitlepage]{revtex4-1}
\usepackage{graphicx}
\usepackage{physics}
\usepackage{float} 
\usepackage{subfigure} 
\usepackage{color}
\usepackage{amsmath}
\usepackage{amsfonts}
\usepackage[normalem]{ulem}
\usepackage{xspace}
\usepackage[dvipsnames]{xcolor}
\usepackage{dcolumn}
\usepackage[breaklinks,colorlinks,linkcolor=blue,citecolor=blue,urlcolor=blue]{hyperref}
\usepackage{lineno}
\usepackage[raggedright]{titlesec}
\usepackage{titlesec}
\titleformat{\section}
  {\normalfont\fontsize{12}{0}\bfseries}{\thesection}{1em}{}
  \titleformat{\subsection}
  {\normalfont\fontsize{10}{0}\bfseries}{\thesubsection}{1em}{}
  
\begin{document}
\title{Recovering dark states by non-Hermiticity }
\author{Qi Zhou}
\affiliation{Department of Physics and Astronomy, Purdue University, West Lafayette, Indiana, 47907}

\date{\today}

\begin{abstract}
Dark states, which are incapable of absorbing and emitting light, have been widely applied in multiple disciplines of physics. However, the existence of dark states relies on certain strict constraints on the system. For instance, in the fundamental $\Lambda$ system, a perturbation breaking the degeneracy between two energy levels may destroy the destructive interference and demolish the dark state. Here, we show that non-Hermiticity can be exploited as a constructive means to restore a dark state.  By compensating for the undesired perturbations, non-Hermiticity produces unidirectional couplings such that the dark state remains decoupled from the rest of the system. Implementing this scheme in many-body systems, flat bands and edge states can be recovered by losses and gains. Further taking into account interactions, a range of novel quantum phases could arise in such non-Hermitian systems. 

\end{abstract}

\maketitle

Dark states are a fundamental concept in quantum physics~\cite{scully1997quantum}. Light-matter interactions may lead to multiple pathways of transitions between atomic levels. Under appropriate conditions, destructive interference arises such that the atom cannot absorb or emit light and occupy certain states and. This textbook result has profound applications in a broad spectrum of topics, ranging from electromagnetically induced transparency and the storage of quantum information to the preparation of novel quantum states~\cite{fleischhauer2005electromagnetically,marangos1998electromagnetically,boller1991observation, diehl2008quantum}. It is worth pointing out that flat bands in condensed matter arise from the same mechanism. 
In certain lattice models, localized orbitals become the eigenstates due to destructive interference between multiple tunneling processes, which leads to vanishing probabilities of occupying lattice sites  other than those in the localized orbitals. Prototypical examples include Lieb lattice, sawtooth lattice, and Crutz lattice, to name a few~\cite{mukherjee2015observation,slot2017experimental,zhang2015one,zhang2014shaping,kang2020topological}. 

To establish a dark state, there are certain strict requirements for the quantum system. For instance, in the $\Lambda$ system, the most fundamental example of quantum systems supporting dark states, the two light fields coupling the first two states, which are denoted by $|\uparrow\rangle$ and $|\downarrow\rangle$ in Fig.1,  with the third one $|3\rangle$ must be at resonance with the corresponding transitions. In other words, in the rotating frame, $|\uparrow\rangle$ and $|\downarrow\rangle$ need to be degenerate. If a perturbation breaks the degeneracy, for instance, an energy difference between $|\uparrow\rangle$ and $|\downarrow\rangle$, the destructive interference is suppressed and the dark state disappears. An important question thus arises: how to re-establish the dark state in the presence of perturbations?

In this work, we show that non-Hermitian couplings, such as losses and gains, can be introduced to the Hamiltonian as an efficient protocol to restore the dark state. This method can be applied to the $\Lambda$ system in the presence of the most generic perturbations. Any undesired couplings between  $|\uparrow\rangle$ and $|\downarrow\rangle$ can be compensated by the corresponding non-Hermitian terms. As such, the dark state interacts with the rest of the system only by unidirectional couplings and thus remains stable. This framework can also be  used in a many-body system to overcome perturbations that destroy flat bands. Whereas there have been considerable efforts to study non-Hermitian flat bands in the literature~\cite{leykam2017flat,ramezani2017non,biesenthal2019experimental, zhang2020non,zhang2020localization,maimaiti2021non,li2024localization}, here, the generic principle of restoring dark states provides us with a systematic means to retain flat bands.  For instance, undesired couplings in a Raman lattice~\cite{zhao2022observing}, which is equivalent to coupled Su-Schrieffer-Heeger chains~\cite{zhang2017two},  lead to finite dispersion of the energy bands.  Adding losses and gains reinstates the flat bands by creating localized orbitals that unidirectionally couple to other lattice sites. These localized orbitals span a few lattice sites such that on-site interactions are sufficient to induce charge-density waves and other novel many-body phases in non-Hermitian systems. 

\begin{figure}
    \centering
    \includegraphics[width=0.45\textwidth]{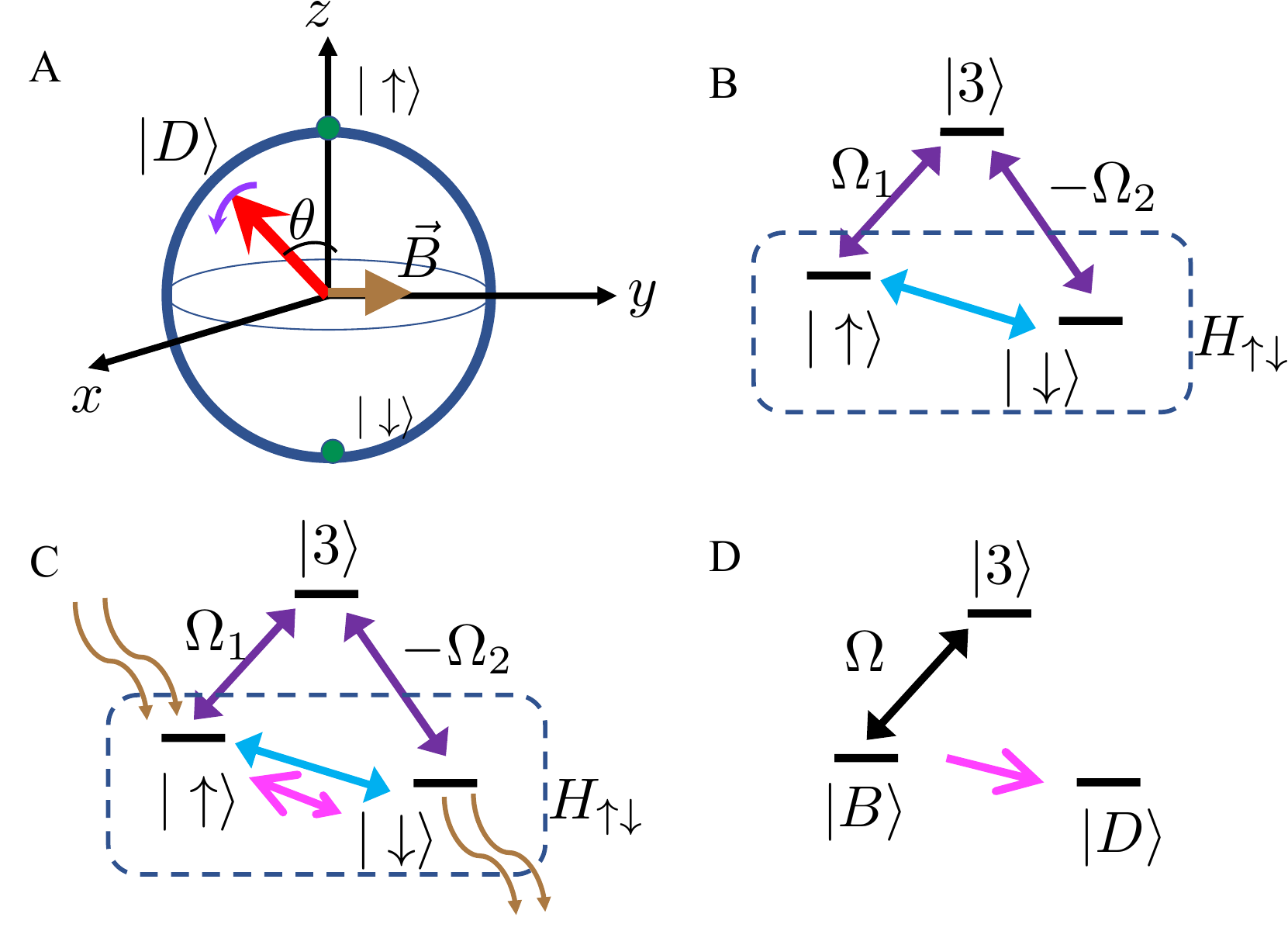}
    \caption{(A) A dark state is represented by a red vector oriented along a certain direction on the Bloch sphere. $\theta$ is determined by $\Omega_1$ and $\Omega_2$. A magnetic field applied to the subspace formed by $|\uparrow\rangle$ and $|\downarrow\rangle$ rotates the spin and $|D\rangle$ is no longer the eigenstate. (B) In a generic three-level system, $H_{\uparrow\downarrow}$ may be finite such that the lowest two states are not degenerate and are also coupled (blue arrow). (C) Adding gain/loss (wiggle vectors) and non-Hermitian couplings (magenta arrows with asymmetric heads) stabilizes $|D\rangle$ as the eigenstate. (D) In the basis of dark and bright states, non-Hermiticity establishes a unidirectional coupling from $|B\rangle$ to $|D\rangle$.  }  
    \label{fig:fig1}
\end{figure}

A $\Lambda$ system is described by a Hamiltonian, 
\begin{equation}
H_0=\Omega_1|\uparrow\rangle\langle 3|-\Omega_2|\downarrow\rangle\langle 3|+h.c.
\end{equation}
Since the phase of the couplings can be gauged away, we have chosen real $\Omega_{1,2}$ to simplify notations. $\hbar=1$ has been chosen in this manuscript. The dark state is written as 
\begin{equation}
|D\rangle=\cos(\theta/2)|\uparrow\rangle+\sin(\theta/2)|\downarrow\rangle, 
\end{equation}
where $\cos(\theta/2)=\Omega_2/\Omega$, $\sin(\theta/2)=\Omega_1/\Omega$, and $\Omega=\sqrt{\Omega_1^2+\Omega_2^2}$. As shown by Fig.(\ref{fig:fig1}A), $|D\rangle$ is denoted by a vector on the Bloch sphere. Using $|D\rangle$ and the orthogonal $|B\rangle=-\sin(\theta/2)|\uparrow\rangle+\cos(\theta/2)|\downarrow\rangle$, the Hamiltonian is rewritten as 
\begin{equation}
H_0=\Omega |B\rangle \langle 3|+h.c.
\end{equation}
$|D\rangle$ disappears from the above equation and thus is invisible to the light fields. This textbook result is built upon the requirement that the Hamiltonian governing the subspace formed by $|\uparrow\rangle$ and $|\downarrow\rangle$ vanishes, $H_{\uparrow\downarrow}=0$. This guarantees that $|D\rangle$ remains stationary and the two pathways to $|3\rangle$, one from $|\uparrow\rangle$ via $|\uparrow\rangle$ and the other from $|\downarrow\rangle$ via $|\downarrow\rangle$, always destructively interfere. 

In reality, $H_{\uparrow\downarrow}$ may not vanish. As shown by Fig.(\ref{fig:fig1}B), the lasers coupling $|\uparrow\rangle$ or $|\downarrow\rangle$ to $|3\rangle$ may be off-resonance such that $|\uparrow\rangle$ and $|\downarrow\rangle$ are not degenerate in the rotating frame. Alternatively, the direct couplings between $|\uparrow\rangle$ and $|\downarrow\rangle$ could be finite. In the most generic case, we write $H_{\uparrow\downarrow}$ as 
\begin{equation}
H_{\uparrow\downarrow}=\vec{B}_R\cdot \vec{S}, \label{Hud}
\end{equation}
where $\vec{B}_R=(B_R^x, B_R^y, B_R^z)$ is a real magnetic field, $\vec{S}=\frac{1}{2}\vec{\sigma}$ and $\vec{\sigma}$ represents the Pauli matrices. Unless $\vec{B}_R$ aligns along the same direction as $|D\rangle$, $H_{\uparrow\downarrow}$ rotates $|D\rangle$ on the Bloch sphere. As such, either the weight of $|\uparrow\rangle$ or the phase of ($\downarrow$ ) changes as time goes by and the interference between the two pathways to $|3\rangle$ is no longer destructive. 
In other words, a generic finite $\vec{B}_R$ destroys the dark state.

To restore the dark state, we consider adding non-Hermitian terms to the Hamiltonian, 
\begin{equation}
H_{\uparrow\downarrow}=\vec{B}\cdot \vec{S}=(\vec{B}_R+i\vec{B}_I)\cdot \vec{S}\label{Hudp},
\end{equation}
where $i\vec{B}_I$ represents an imaginary magnetic field. As shown in Fig.(\ref{fig:fig1}C), $B^z_I$ characterizes the gain and loss in $|\uparrow\rangle$ and $|\downarrow\rangle$, respectively. $B^x_I$ and $B^y_I$ give rise to non-reciprocal couplings between $|\uparrow\rangle$ and $|\downarrow\rangle$. Such a spin is thus subject to a complex magnetic field $\vec{B}=\vec{B}_R+i\vec{B}_I$, which has been realized in a variety of systems. To see that $i\vec{B}_I$ can compensate $\vec{B}_R$ and make $|D\rangle$ stationary, we go to the basis formed by $|D\rangle$ and $|B\rangle$. We will see that $i\vec{B}_I$ leads to unidirectional couplings between $|D\rangle$ and $|B\rangle$.  

 Define a unitary transform $U=e^{i\theta Sy}$ such that $US_{z'} U^{-1}=S_z$, where $S_{z'}=\cos(\theta)S_z+\sin(\theta)S_x$ is the projection of $\vec{S}$ in the quantization axis along the direction of $|D\rangle$. In this new basis, $H_{\uparrow\downarrow}-  {H}'_{\uparrow\downarrow}=U{H}_{\uparrow\downarrow}U^{-1}=\vec{{B}}'\cdot \vec{S}$. A straightforward calculation show that 
\begin{align}
&{B'}^x=B^x\cos\theta-B^z\sin\theta,\nonumber\\
&{B'}^y=B^y,\nonumber\\
&{B'}^z=B^x\sin\theta+B^z\cos\theta,\label{Bp}
\end{align}
where each component, ${B}'^{x,y,z}$,  is complex.  Viewing $|D\rangle$ and $|B\rangle$ as a two-level system, $H'_{\uparrow\downarrow}$ can be rewritten as 
\begin{equation}
H'_{\uparrow\downarrow}=\frac{\delta}{2}|D\rangle\langle D|-\frac{\delta}{2}|B\rangle\langle B|+t_L|D\rangle\langle B|+t_R|B\rangle\langle D|, \label{H2}
\end{equation}
where 
\begin{align}
&t_L=B_R'^x+B'^y_I+i(B'^x_I-B'^y_R), \nonumber\\
&t_R=B_R'^x-B'^y_I+i(B'^x_I+B'^y_R),\nonumber\\
&\delta=B'^z_R+iB'^z_I.
\end{align}
A complex $\vec{B}$, or equivalently a complex $\vec{B}'$, means that all the coefficients in Eq.(\ref{H2}), in general, are complex and $t_L\neq t_R^*$.

To guarantee that $|D\rangle$ remains a stationary state under the Hamiltonian $H'_{\uparrow\downarrow}$, it is required that  $t_L=\text{Im}\delta=0$.
A vanishing $t_L$ means that the coupling between $|D\rangle$ and $|B\rangle$ is unidirectional, only from $|B\rangle$ to $|D\rangle$, not vice versa. As such, $|D\rangle$ remains an eigenstate of $H'_{\uparrow\downarrow}$.  A vanishing $\text{Im} \delta$ guarantees that the energy shift of $|D\rangle$ and $|B\rangle$ due to $\vec{B}'$ is real.  Satisfying both these two criteria, $|D\rangle$ remains a stationary state of  $H'_{\uparrow\downarrow}$. We thus obtain,
$ B'^x_I=B'^y_R$, $B'^y_I=-B'^x_R$, $B'^z_I=0$,
which allows us to fix $\vec{B}'_I$ using $\vec{B}'_R$. We further apply the inverse transform of Eq.(\ref{Bp}), and then relationship between $\vec{B}_R$ and $\vec{B}_I$ is obtained,
\begin{align}
&B^x_I=- B^y_R\cos\theta,\nonumber\\
&B^y_I=B^x_R\cos\theta-B^z_R\sin\theta,\nonumber\\
&B^z_I=B^y_R\sin\theta.
\end{align}


The above set of equations has a clear physical meaning. When the $y$-component of the real magnetic field $B^y_R$ is finite, $|D\rangle$ tends to rotate about the $y$ axis, as shown in Fig.(\ref{fig:fig1}). Consequently, $\langle S^z\rangle$ ($\langle S^x\rangle$) tends to decrease (increase). To compensate for the increase of $\langle S^z\rangle$, an imaginary magnetic field $iB^z_I$ should be added in the $z$-direction, resulting in a gain in $|\uparrow\rangle$ and a loss in $|\downarrow\rangle$. Similarly,  an imaginary magnetic field $iB^x_I$ is required to cancel the increase of $\langle S^x\rangle$. As such, $|D\rangle$ remains stationary due to the interplay between a real magnetic field in the $y$ direction and an imaginary field having finite projections in both the $x$ and $z$ directions. Similar discussions apply to cases of finite $B^x_R$ and $B^{z}_R$, which can also be compensated by corresponding imaginary magnetic fields. 

\begin{figure}[t]
    \centering
    \includegraphics[width=0.5\textwidth]{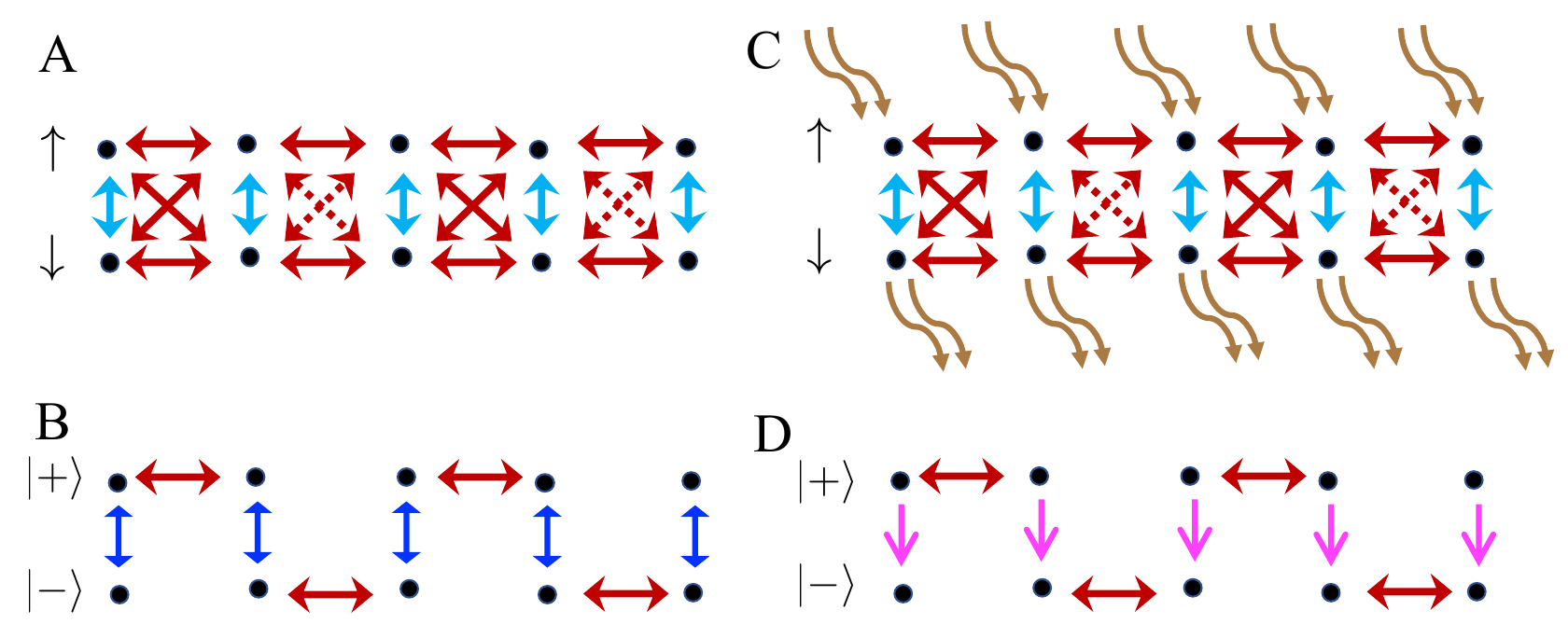}
    \caption{(A) A generalized Crutz ladder. Dashed arrows denote the alternating sign change of the intersite spin flipping term. (B) Rewriting the generalized Crutz ladder in the basis of $|+\rangle_n$ and $|-\rangle_n$ leads to a two-leg SSH chain with glide reflection symmetry. (C) Adding gain and losses to the generalized Crutz ladder is equivalent to a two-leg SSH chain coupled by unidirectional hoppings in (D).}  
    \label{fig:fig2}
\end{figure}

The above discussions of $\Lambda$ systems lay the foundation for using non-Hermiticity to restore dark states in both few-body and many-body systems. 
Here, we apply this protocol to study flat band physics, which is of particular interest in many-body physics~\cite{mukherjee2015observation,slot2017experimental,zhang2015one,zhang2014shaping,kang2020topological}. Despite the fact that the very rich models of flat bands appear different from each other, the essential physics is the same. In the lattice model, $|\uparrow\rangle$ and $|\downarrow\rangle$ in Fig.(\ref{fig:fig1}) correspond to two lattice sites in the localized orbital. $\Omega_1$ and $\Omega_2$ are now the tunneling to a site $|3\rangle$ outside the localized orbital. Our scheme applies to generic perturbations that demolish
the destructive interference between the two pathways to $|3\rangle$ and thus restore the flat bands. 

As a concrete example, we consider a lattice model as shown by Fig.(\ref{fig:fig2}B),  
\begin{align}
H_L&=\sum_{n\sigma} (\frac{t}{2} a^\dagger_{n\sigma}a_{n+1\sigma}+h.c.)+\sum_{n} (-1)^n(\frac{t}{2}a^\dagger_{n\uparrow}a_{n+1\downarrow}+h.c.)\nonumber\\
+&\sum_n i\Gamma (-a^\dagger_{n\uparrow} a_{n\uparrow}+a^\dagger_{n\downarrow} a_{n\downarrow})+((\Omega_x-i\Omega_y)a^\dagger_{n\uparrow} a_{n\downarrow}+h.c.),\label{HLa}
\end{align}
where $a^\dagger_{n\sigma}$ is the creation operator at site $n$ for spin-$\sigma$, and $\sigma=\uparrow, \downarrow$. $t/2$ represents the tunneling strength,  
$\Omega_{x,y}$ denotes the onsite spin flip strength, and $\Gamma$ is the strength of the loss and gain. As shown by Fig.(\ref{fig:fig2}A), when $\Gamma=\Omega_x=\Omega_y=0$, $H_L$ could be regarded as a generalized Crutz ladder, where the factor $(-1)^n$ in $H_L$ leads to a unit cell composed of four lattice sites~\cite{zhang2017two}. It can be realized by applying a Raman lattice to atoms with two hyperfine spin states~\cite{zhang2017two,zhao2022observing}.  A finite $\Gamma$ can be introduced by adding pumping to spin-up and dissipation to spin-down. An extra uniform coupling, such as rf coupling with an appropriately chosen phase relative to the Raman coupling, gives rise to a finite $\Omega_x$ and $\Omega_y$.  A transformation $\tilde{a}^\dagger_{n\uparrow}=(-1)^n{a}^\dagger_{n\uparrow}$ halves the unit cell and $H_L$ describes two coupled bands with different parity. This model can be realized by periodically driving the lattice using two frequencies~\cite{zhang2014shaping,kang2020topological}. \begin{figure}[t]
    \centering
    \includegraphics[width=0.45\textwidth]{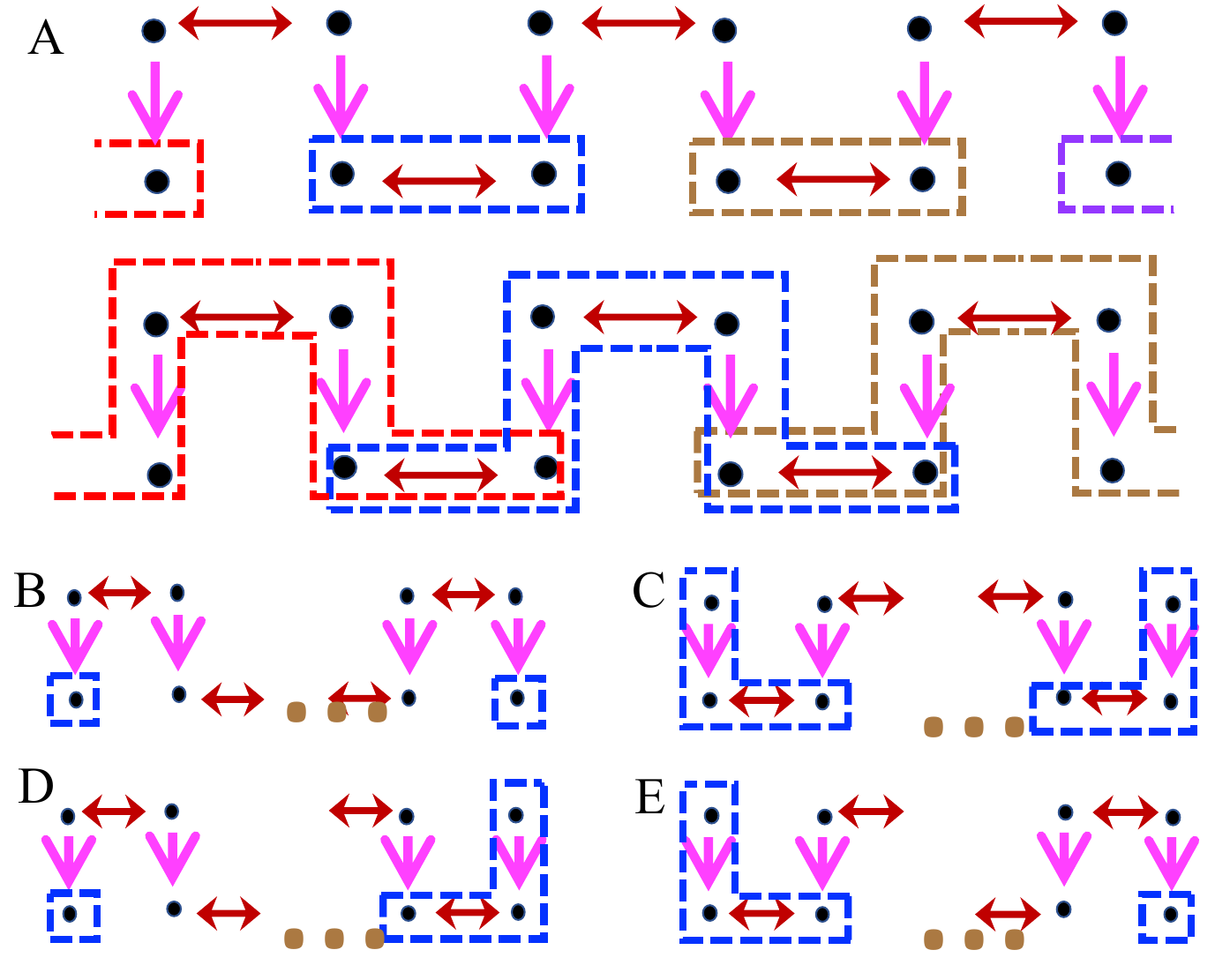}
    \caption{(A) Each state in the first set of localized orbitals consists of two lattice sites in the lower chain. Different colors are used to distinguish orbitals at different locations. Another set of localized orbitals, each of which consists of two sites in the upper chain and four sites in the lower chain. (B-E) Blue boxes highlight the locations of edge states, which consist of either one or three lattice sites. Depending on whether the number of lattice sites is even or odd, there are four possibilities to arrange localized states at edges. }  
    \label{fig:fig3}
\end{figure}

The model in Eq.(\ref{HLa}) was recently studied in the context of non-Hermitian skin effects~\cite{zhou2022engineering}. Here, we consider different physics regarding flat band and edge states. When $\Gamma=\Omega_x=\Omega_y=0$, it has been shown that flat bands arise since the eigenstates of $H_L$ are localized orbitals due to destructive interference between two pathways, one from the spin conserved coupling and the other from spin flipped couplings~\cite{zhang2014shaping}. For more generic cases,  it is more convenient to switch to a different basis $|+\rangle_n=\frac{1}{\sqrt{2}}(|\uparrow\rangle_n+|\downarrow\rangle_n)$, $|-\rangle=\frac{1}{\sqrt{2}}(|\uparrow\rangle_n-|\downarrow\rangle_n)$, where the localized orbitals can be best visualized as that in Fig.(\ref{fig:fig1}D). We define $b^\dagger_{n\uparrow}=(a^\dagger_{n\uparrow}+ a^\dagger_{n\downarrow})/\sqrt{2}$, $b^\dagger_{n\downarrow}=(a^\dagger_{n\uparrow}-a^\dagger_{n\downarrow})/\sqrt{2}$, and we obtain $H_L=\sum_n H_n$, where

\begin{align}
H_n&= t (b^\dagger_{2n\uparrow}a_{2n+1\uparrow}+b^\dagger_{2n+1\downarrow}b_{2n+2\downarrow}+h.c.)+\Omega_x( b^\dagger_{n\uparrow}b_{n\uparrow}\nonumber\\&-b^\dagger_{n\downarrow}b_{n\downarrow})
+(-i\Gamma-i\Omega_y) b^\dagger_{n\uparrow} b_{n\downarrow}+(-i\Gamma+i\Omega_y) b^\dagger_{n\downarrow} b_{n\uparrow}.
\end{align} 
This model describes two chains coupled by asymmetric hoppings $t_{\uparrow}=-i\Gamma-i\Omega_y$ and $t_{\downarrow}=-i\Gamma+i\Omega_y$.  As shown by Fig.(\ref{fig:fig2}B), when $\Omega_x=\Gamma=0$, $H_L$ describes a two-leg SSH ladder with glide reflection symmetry~\cite{zhang2017two}. Alternatively, this model can also be viewed as an SSH chain with long-range couplings or enlarged unit cells\cite{li2018extended,zurita2021tunable,dias2022long,ares2022symmetry,qian2023observation,zhou2023exploring,ara2023topological,mandal2024topological,ghuneim2024topological,mandal2024topological}. For convenience, we will consider it as a ladder formed by two SSH chains. It is apparent that flat bands arise when $\Omega_y=0$.  In the bulk, each eigenstate is composed of two sites in each chain. At the boundary, an isolated site in a single chain is decoupled from the bulk and gives rise to a zero-energy edge state.  In the presence of a finite $\Omega_y$, these two SSH chains are coupled and energy bands are no longer flat.

\begin{figure*}
    \centering
    \includegraphics[width=1\textwidth]{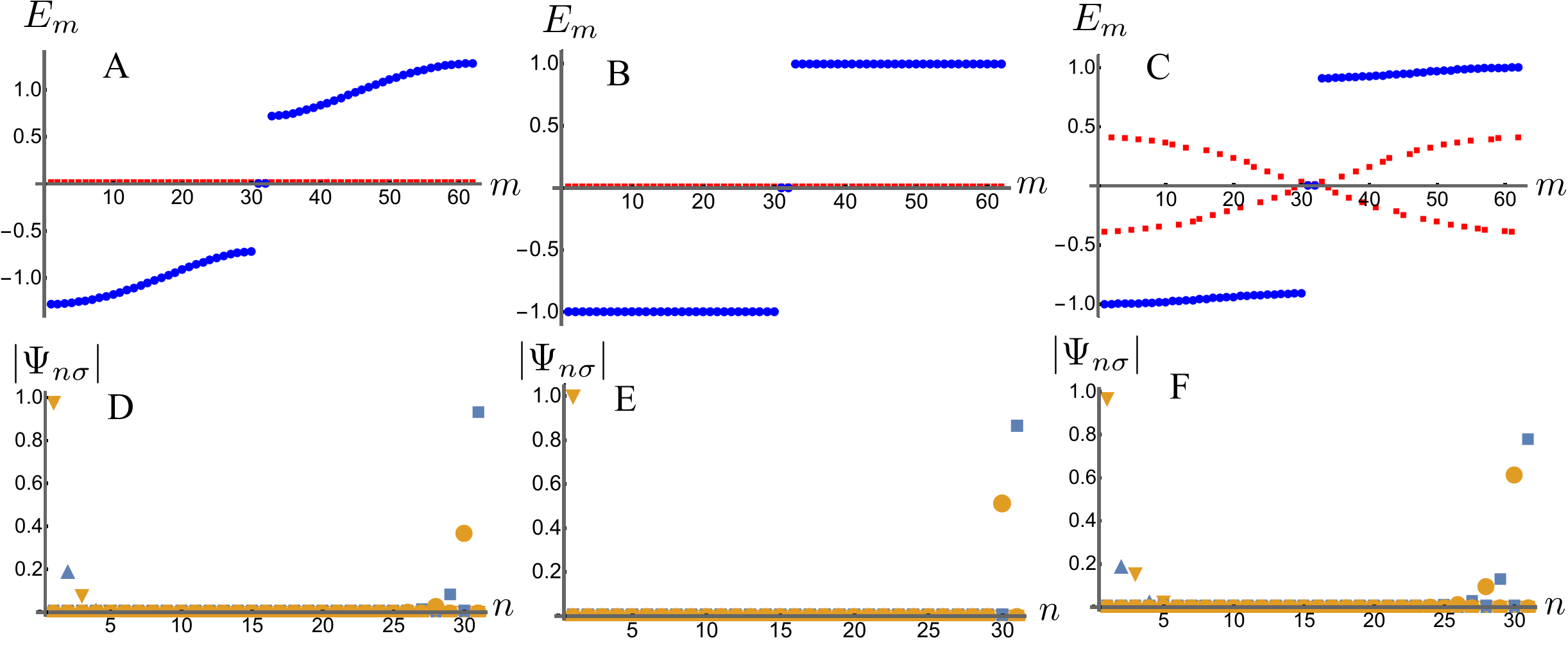}
    \caption{(A-C). The real part (blue dots) and the imaginary part (red boxes) of the energy. $t$ has been chosen as the energy scale, $\Omega_y=0.3$, $\Omega_x=0$. $\Gamma=-0.1, -0.3, -0.5$ in A, B, C, respectively. (D-F) The wavefunctions of the corresponding edge states. Blue up triangles and brown down triangles represent $|\Psi_{n\uparrow}|^2$ and  $|\Psi_{n\downarrow}|^2$ of the left edge state. Blue boxes and brown dots represent $|\Psi_{n\uparrow}|$ and  $|\Psi_{n\downarrow}|$ of the right edge state.  }  
    \label{fig:fig4}
\end{figure*}
To restore the flat bands, a finite $\Gamma=\pm \Omega_y$ can be chosen such that the two SSH chains are coupled by unidirectional hoppings. For instance, when $\Gamma=-\Omega_y$, a particle can hop from the upper chain to the lower one, not vice versa, as shown by Fig.(\ref{fig:fig2}D).  As such, the eigenstates in the bulk are all localized ones. As shown in Fig.(\ref{fig:fig3}), there are two sets of such localized orbitals. To see their origin, we consider a local Hamiltonian 
 in the basis formed by $|-\rangle_{n-2}$, $|-\rangle_{n-1}$, $|+\rangle_{n}$, $|+\rangle_{n+1}$, $|-\rangle_{n+2}$, $|-\rangle_{n+3}$,
\begin{equation}
h=\left(    \begin{array}{cccccc}
    -\Omega_x & t & 0 & 0 & 0& 0 \\ 
    t &  -\Omega_x & -2i \Gamma & 0 & 0 & 0 \\ 
  0 &   0& \Omega_x & t & 0 & 0 \\ 
    0 & 0 & t & \Omega_x& 0 & 0 \\ 
    0 & 0 & 0 & -2i \Gamma & -\Omega_x& t \\ 
    0 & 0& 0 & 0 & t & -\Omega_x\\ 
  \end{array}\right).
\end{equation}
As shown by Fig.(\ref{fig:fig2}D), a particle on any site in the lower chain cannot tunnel to the upper chain. As such, the subspace formed by the above six basis states is decoupled from the rest of the system.

We denote the eigenstates of $h$ as $\phi_i$, where $i=1,2,3,4,+,-$, i.e., $h\phi_i=E_i\phi_i$. Four of such eigenstates have finite occupancy only at the lower chain, $\phi_1=(1, 1, 0,0,0,0)^T/\sqrt{2}$, $\phi_2=(1, -1, 0,0,0,0)^T/\sqrt{2}$, $\phi_3=(0,0,0,0,1,1)^T/\sqrt{2}$, $\phi_4=(0,0,0,0,1,-1)^T/\sqrt{2}$, where $T$ denotes the transpose of a matrix. These states provide us with the first set of localized eigenstates shown in Fig.(\ref{fig:fig3}A),
 \begin{equation} |D^{\pm}\rangle_{n\downarrow }=D^\dagger_{n\downarrow\pm}|0\rangle \end{equation} 
 where $D^\dagger_{n\downarrow\pm}= \frac{1}{\sqrt{2}}(b^\dagger_{n\downarrow}\pm b^\dagger_{n+1\downarrow})$. The energy is given by 
 \begin{equation}
 E_{\downarrow\pm}=-\Omega_x\pm t.\label{Ed}
 \end{equation}
 Apparently, $|D^\pm \rangle_{n\downarrow}$ is decoupled from the rest of the system.  The other two eigenstates, $\phi_\pm$, have finite occupancy in both chains, providing the other set of localized orbitals of $H_L$, 
\begin{equation}
|D^{\pm}\rangle_{n\uparrow }=D^\dagger_{n\uparrow\pm}|0\rangle,
\end{equation}
where $D^\dagger_{n\uparrow\pm}=\phi^T_\pm\cdot \mathcal B$ and $\mathcal B$ is a short-hand notation,  $\mathcal B=( b^\dagger_{n-2\downarrow}, b^\dagger_{n-1\downarrow}, b^\dagger_{n\uparrow}, b^\dagger_{n+1\uparrow}, b^\dagger_{n+2\downarrow}, b^\dagger_{n+3\downarrow})^T$. The eigenenergies are 
\begin{equation}
E_{\uparrow\pm}=\Omega_x\pm t.\label{Eu}
\end{equation}
We thus conclude that there are four flat bands in the bulk, whose energies are written in Eq.(\ref{Ed},\ref{Eu}). When $\Omega_x=0$, the number of flat bands reduces to two, and each band is doubly degenerate.

In addition to the localized orbitals in the bulk, it is also interesting to investigate the edge states. As shown by Fig.(\ref{fig:fig3}B-E), there are two types of localized states at the edges. One consists of a single lattice site at either end of the lower chain. Due to the unidirectional couplings between the up and lower chains, such states are apparently isolated from other lattice sites and the energies are given by $E_{e\downarrow}=-\Omega_x$. Another type of edges state are formed by three lattice sites. For instance, the edge state at the right end of the lattice with odd number of lattice sites arises from the Hamiltonian, 
\begin{equation}
h_{e\uparrow}=\left(    \begin{array}{ccc}
    \Omega_x &0 & 0  \\ 
    -2i\Gamma &  -\Omega_x & t\\ 
  0 &   t& -\Omega_x 
        \end{array}\right),
\end{equation}
where the basis states are $|+\rangle_L$, $|-\rangle_L$, and $|-\rangle_{L-1}$. $L$ is the number of lattice sites. We denote the eigenstates of $h_{e\uparrow}$ as $\tilde{\phi}_i$, where $i=1,2,3$. Two of the three eigenstates are $(0,1,1)^T/\sqrt{2}$, and $(0,1,-1)^T/\sqrt{2}$. They coincide with the localized states in the bulk, i.e., $|D^{\pm}\rangle_{L-1}$. The third state could be expressed as 
\begin{equation}
|E^{\uparrow}\rangle_{n\uparrow }=\tilde{\phi}^T_3\cdot \mathcal{\tilde{B}} |0\rangle,
\end{equation}
where $\mathcal{\tilde{B}} =(b^\dagger_{L\uparrow}, b^\dagger_{L\downarrow}, b^\dagger_{L-1\uparrow})^T$. Its energy is given by $E_{e\uparrow}=\Omega_x$.

Whereas we have been focusing on $\Gamma=|\Omega_y|$, it is useful to discuss more generic scenarios. When $\Gamma\neq |\Omega_y|$, the coupling between the upper and lower chains are no longer unidirectional and the previously discussed localized orbitals are going to be coupled to each other. Nonetheless, the bandwidth is small when either $t_{\uparrow}$ or $t_{\downarrow}$ is small. In particular, the energy gap between the bulk states and the edge state may remain finite before crossing a critical point, as shown in Fig.(\ref{fig:fig4}). If we consider $\Gamma>-\Omega_y$, the wavefunction of the zero-energy edge states at the left end is written as
\begin{equation}
|L\rangle=\sum_{n\sigma}\Psi_{n\sigma}b^\dagger_{n\sigma}|0\rangle. 
\end{equation}
Using the zero energy solution to the Schrodinger equation, we find that $\Psi_{n\sigma}$ satisfies 
\begin{align}
&\Psi_{n\uparrow}=0, \,\,\, n\in even,\\
&\Psi_{n\downarrow}=0,  \,\,\, n\in odd,
\end{align}
and 
\begin{align}
&\Psi_{2n+1\uparrow}= \left(\frac{t_\uparrow t_\downarrow}{t^2}\right)^n \frac{t_{\uparrow}}{t}\Psi_{0\downarrow}, \\
&\Psi_{2n\downarrow}=\left(\frac{t_\uparrow t_\downarrow}{t^2}\right)^n \Psi_{0\downarrow}. 
\end{align}
We thus see that the wavefunction decays exponentially with the localization length $\sigma=\ln^{-1}( \frac{t^2}{|t_\uparrow t_\downarrow|})$.

Similar discussions apply to the edge state at the right end, whose wavefunction exponentially decays with the same localization length. We thus conclude that the critical point is determined by $t_\uparrow t_\downarrow=t^2$. When $\Gamma$ continuously decreases down to $-\Omega_y$, $t_{\uparrow}$ vanishes and the localization length becomes zero, as previously discussed. Further decreasing $\Gamma$, $t_{\uparrow}$ changes sign and the energy of a bulk state acquires a finite imaginary part. Nonetheless, the energy of the edge state remains zero. Only when $|t_{\uparrow}t_{\downarrow}|$ becomes larger again, the edge states disappear and any energy eigenstate has a finite imaginary part of the energy. 

Since $t_{\uparrow}$ and $t_{\downarrow}$ depend on $\Gamma$ and $\Omega_y$, it is worth highlighting a case of particular interest. Consider $\Omega_y>t$, when $\Gamma=0$, the model reduces to an ordinary SSH chain in the regime without an edge state, distinct from the case where the edge states exist if $\Omega_y<t$. Turning on a finite $\Gamma$, $|t_{\downarrow}t_{\uparrow}|=\Omega_y^2-\Gamma^2$ decreases. At the critical point, $\Omega_y^2-\Gamma^2=t^2$, and the edge states appear. This means that non-Hermiticity could be implemented as a constructive means to recover the edge state and turn a trivial phase to a nontrivial one. 

Whereas we have considered a vanishing $\Omega_y$, similar discussions apply to a finite $\Omega_y$. Despite the fact that there are quantitative differences such as finite energies of the edge states and two degenerate bands in the bulk splitting into four bands, the essential physics of the existence of edge states remains the same. 

In the above discussions, we have been focusing on non-interacting systems. It is useful to comment on the interaction effects. Since each of the localized orbitals in the flat band limit occupies multiple sites, on-site interactions become sufficient to induce novel quantum many-body states. When we consider a repulsive onsite interaction, 
\begin{equation}
H_{int}=U\sum_n b^\dagger_{n\uparrow}b_{n\uparrow}b^\dagger_{n\downarrow}b_{n\downarrow},
\end{equation}
where $U>0$, occupying overlapping localized orbitals in Fig.(\ref{fig:fig3}) leads to an energy penalty. Such an energy penalty gives rise to certain novel quantum phases at the ground state. For instance, when $\Omega_x<0$ and $0<t<|\Omega_x|$, the energy gap between the lowest two bands is written as $\Delta=E_{\uparrow+}-E_{\uparrow-}=2t$. When $U\ll t$, it is sufficient to project $H_{int}$ to the lowest band. At filling factor of $1/8$, the lowest band is half filled. Due to the vanishing bandwidth, bosons do not condense and only one of the two successive localized orbitals, $|D^-\rangle_{n\uparrow}$, is occupied to avoid the energy penalty. As such, 
 a charge density wave arises as the ground state, and the wavefunction is written as 
\begin{equation}
|G\rangle=\prod_{n\in even}D^\dagger_{n\uparrow+}|0\rangle. 
\end{equation} 
or $\prod_{n\in odd}D^\dagger_{n\uparrow+}|0\rangle$. Doping this charge density wave, overlapping localized orbitals lead to density assisted tunneling and pairing hoppings and a supersolid or paired superfluid may appear~\cite{huber2010bose,takayoshi2013phase, tovmasyan2013geometry}. 

We have shown that non-Hermitian Hamiltonians could be implemented to recover the dark states. Applying this scheme to lattice systems could create or recover localized eigenstates, including flat bands in the bulk and edge states at the boundaries. We hope that our work will stimulate more interest in studying such non-Hermiticity restored dark states and their applications in many-body physics.






\begin{acknowledgments}
This work is supported by the Air Force Office of Scientific Research under award number FA9550-23-1-0491.
\end{acknowledgments}

\bibliographystyle{apstest}
\bibliography{dc.bib}

\end{document}